\documentclass[english,onecolumn,showpacs]{revtex4}
\usepackage[latin1]{inputenc}
\usepackage{amsmath}
\usepackage{graphicx}
\usepackage[english]{babel}

\batchmode

\newcommand{\beq}{\begin{equation}}
\newcommand{\eeq}{\end{equation}}
\newcommand{\ket}[1]{| #1 \rangle}
\newcommand{\bra}[1]{\langle #1 |}

\usepackage{babel}

\begin{document}

\title{Optimal Irreversible Stimulated Emission}

\author{D. Valente$^{1}$}\email{valente.daniel@gmail.com}
\author{Y. Li$^{2}$}
\author{J. P. Poizat$^{1}$}
\author{J. M. G\'erard$^{3}$}
\author{L. C. Kwek$^{2}$}
\author{M. F. Santos $^{4}$}
\author{A. Auff\`eves$^{1}$}


\affiliation{$^{1}$ Institut N\'eel-CNRS, Grenoble, France}

\affiliation{$^{2}$ Centre for Quantum Technologies, National Universtity of Singapore}

\affiliation{$^{3}$ CEA/INAC/SP2M, Grenoble, France}

\affiliation{$^{4}$ Departamento de F\'isica, Universidade Federal de Minas Gerais, Belo Horizonte, Brazil}

\begin{abstract}
We study the dynamics of an initially inverted atom in a semi-infinite waveguide, in the presence of a single propagating photon. We show that atomic
relaxation is enhanced by a factor 2, leading to maximal bunching in the output field. This optimal irreversible stimulated emission is a novel phenomenon that can be observed with state of the art solid state
atoms and waveguides. When the atom interacts with two one-dimensional electromagnetic environments, preferential emission in the stimulated field can be exploited to efficiently amplify a classical or a quantum state.

\end{abstract}

\maketitle

\section{Introduction}
\hspace{0.3cm}
Stimulated emission is a central concept of quantum optics. Initially introduced by Einstein \cite{Einstein} in the case of a macroscopic number of emitters and electromagnetic modes, it describes the increase of light-matter coupling
with the number of photons in the mode. Because of this non-linear behavior, an initially inverted gain medium preferentially emits light in the stimulated mode,
a feature that led to the development of lasers and amplifiers technology \cite{siegman}.
Latter experimental developments have allowed to design optical media sensitive at the single photon level. One paradigmatic example is given by a two-level atom
strongly coupled to a monomode cavity \cite{haroche,kimble}. In this specific case, spontaneous emission is reversible and gives rise to the so-called vacuum Rabi oscillation, whereas the presence of a single stimulating photon enhances the frequency of the oscillation by a factor $\sqrt{2}$ \cite{haroche2}.
More recently, another class of giant optical non-linear medium has emerged, like atoms weakly coupled to directional cavities \cite{1D,GNL}, superconducting qubits in circuit QED \cite{astafiev1D}, or quantum dots in photonic wires  \cite{jclaudon}.  Such atoms that interact with only one direction of the light field open promising perspectives in quantum communication and information processing, as they are expected to provide efficient single photon transistors \cite{astafiev,hwang} or photonic gates \cite{Abdumalikov,Hoi}.
These one-dimensional systems evidence new fundamental effects which have been extensively studied both experimentally \cite{lukin07, vetsch} and theoretically \cite{hakutamatata,shenfan,gritsev}, just to mention a few examples.

Here we explore the dynamics of stimulated emission for a two-level atom embedded in a semi-infinite waveguide. We show that a single resonant photon, properly shaped, propagating in the waveguide can shorten the atomic lifetime by a factor 2, leading to significant bunching in the output light field. 
Such optimal irreversible stimulated emission is a new phenomenon: in the context of cavity quantum electrodynamics, stimulation by a single-photon is either optimal, but reversible or irreversible, but not optimal. Building on this effect, we revisit the case of an atom interacting with two one-dimensional electromagnetic fields, as it is the case for a two-level atom in a transmitting waveguide, or a lambda shaped three level atom in a half waveguide.
Under optimal conditions, light emission is twice more probable in the stimulated field than in the field prepared in the vacuum state. This effect is discussed in the context of amplification and quantum cloning.

\section{Model and methods}
\begin{figure}[h!]
\begin{center}
\includegraphics[width=0.5\linewidth]{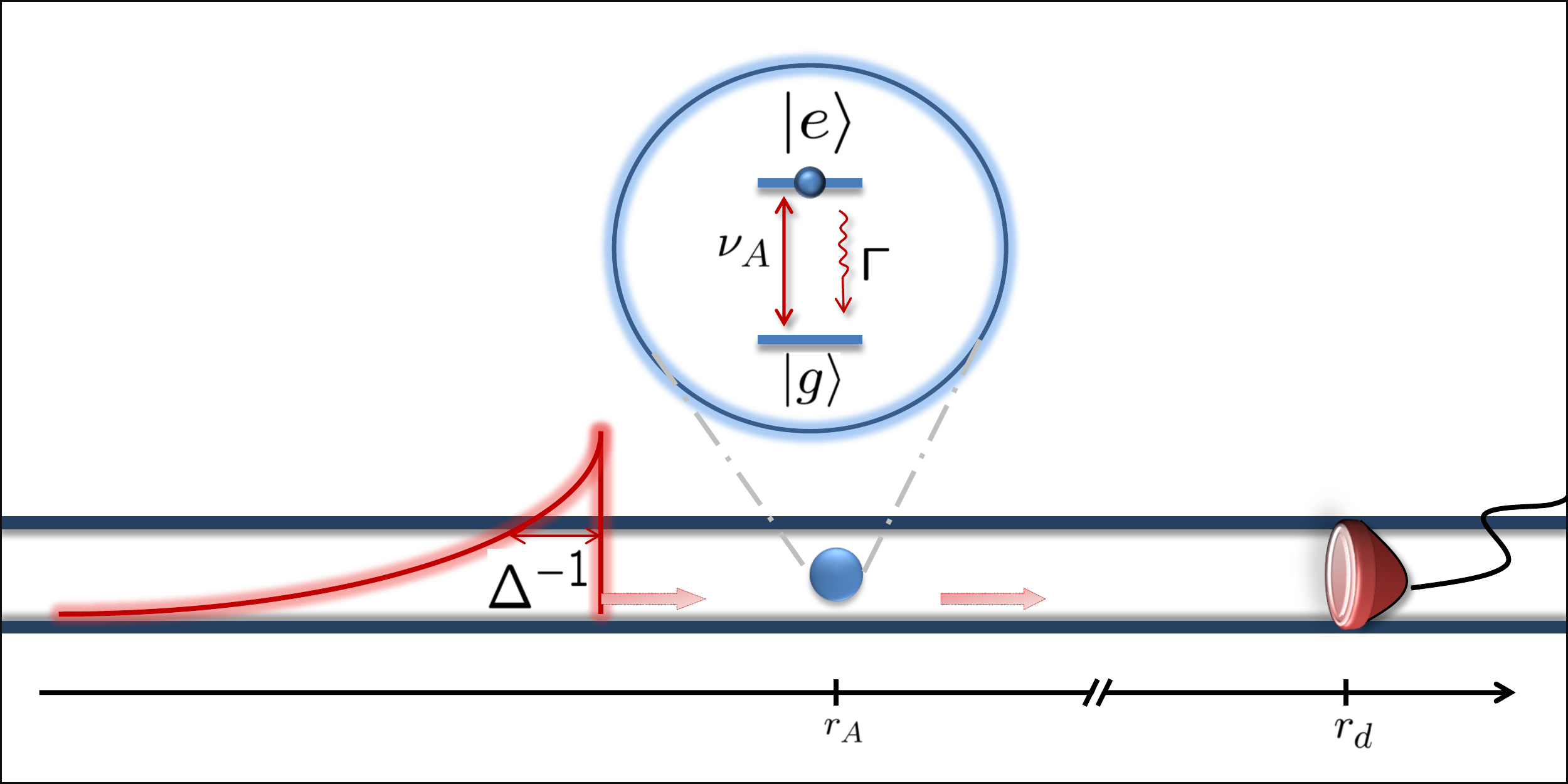}
\caption{Scheme of the two-level atom stimulated by a single photon pulse of linewidth $\Delta$. The spontaneous decay rate is $\Gamma$ and the transition frequency is $\nu_A$. The direction indicated by the arrows is the only allowed one, which models the presence of a mirror close to the atom, at position $r_A$. The light field is detected by a photodetector positioned at $r_d$, put sufficiently far from the atom.}
\label{fig:1datom}
\end{center}
\end{figure}

The system under study is represented in Fig.\ref{fig:1datom}. An initially inverted atom is embedded in a waveguide where only one direction of propagation is allowed, which could be realized by a semi-infinite waveguide, closed by a perfect mirror. Atomic emission
may be stimulated by a single photon pulse propagating in the waveguide. The total Hamiltonian is given by

\beq
H = H_{\mathrm{1D}}+H_{\mathrm{atom}}+H_{\mathrm{int}},
\eeq
which is the same system analyzed in Ref. \cite{kojima}. It describes an emitter interacting with a one-dimensional electromagnetic field,
given by the Hamiltonian $H_{\mathrm{1D}} = \sum_{\nu=0}^\infty\hbar\nu\ a_\nu^\dagger a_\nu $, where $\nu$ is the frequency of the $a_\nu$ mode \cite{kojima}.
The atom with frequency $\nu_A$ is modeled by $H_{\mathrm{atom}} = \hbar \nu_A\ \sigma_+ \sigma_-$, where $\sigma_- = |g\rangle\langle e|$ and
$\sigma_+ = \sigma_-^\dagger$. The dipole interaction is given in the rotating-wave approximation by
\beq
H_{\mathrm{int}} = -i\sum_{\nu=0}^\infty \hbar g_\nu \left[a_\nu \sigma_+ - \mathrm{H.c.} \right]
\label{semi1D}\eeq
for the atom centered at the origin, where $g_\nu$ is the coupling strength.

The solution of the problem is Hamiltonian, so the total number of excitations is conserved. An Ansatz is constructed for the complete wavefunction $|\psi(t)\rangle$,
\beq
\ket{\psi(t)} =\sum_{\nu=0}^\infty \psi_\nu(t)\ a_\nu^\dagger  \ket{e,0}
+\sum_{\nu,\nu'=0}^\infty \phi_{\nu,\nu'}(t)\ a_\nu^\dagger a_{\nu'}^\dagger \ket{g,0},
\eeq
that evolves according to the Schroedinger equation,
\beq
\frac{\partial}{\partial t} \psi_\nu = -i(\nu_A + \nu) \psi_\nu - 2\sum_{\nu'=0}^\infty g_{\nu'} \phi_{\nu,\nu'},
\label{psinu}\eeq
and
\beq
\frac{\partial}{\partial t} \phi_{\nu,\nu'} = -i(\nu+\nu')\phi_{\nu,\nu'} + \frac{1}{2}(g_{\nu'} \psi_{\nu}+g_{\nu} \psi_{\nu'}).
\label{phinm}\eeq

\subsection{Excited-state amplitude in real-space representation}
For the following analysis we adopt a real space representation of the quantum state,
$\psi(r,t) \equiv \sum_{\nu = 0}^\infty \psi_\nu(t) \ e^{ik_\nu r}$, where
$k_{\nu}= |\vec{k}_{\nu}| = \nu/c$.
With the help of the transformation $\sum_\nu \nu \ \psi_\nu e^{ik_\nu r} = -ic\partial_r\sum_\nu \psi_\nu e^{ik_\nu r}$,
we rewrite Eq. (\ref{psinu}) in the form
$\left[\frac{\partial}{\partial t} + c\frac{\partial}{\partial r}\right]\psi(r,t) = -i\nu_A\ \psi(r,t)$
$- 2 \sum_{\nu,\nu'} g_{\nu'}\ \phi_{\nu,\nu'} e^{ik_\nu r}$.
In the trivial uncoupled case $g_\nu = 0$, for instance,
the solution to the equation above is simply a product of a free time-dependent atomic evolution $f(t) = \exp(-i\nu_A t)$ and a propagating
pulse $p(r - ct)$, where $p(r)$ is the initial pulse wavefunction.
In this paper we integrate Eq.(\ref{phinm}) choosing the initial condition $\phi_{\nu,\nu'}(0) = 0$,
so that the atom is initially in the excited state with a single-photon propagating in its direction.
As usual, we assume the fast rotating reference frame,
$\psi(r,t) = \tilde{\psi}(r,t)\exp(-i\nu_A t)\exp(-i\nu_L(t- r/c))$, where $\nu_L$ is the central frequency of the incident wavepacket.
Two terms in our equations deserve further considerations. The first is $\int_0^t dt' \tilde{\psi}(r- c(t-t'),t')$
$\times \sum_{\nu'} g_{\nu'}^2 e^{-i(\nu'-\nu_A)(t-t')}$ which equals to $(\Gamma/2)\ \tilde{\psi}(r,t)$, under Markovian approximation.
The decay constant is $\Gamma \equiv \sum_{\nu'}2\pi g^2_{\nu'} \delta(\nu'-\nu_A)$, as obtained by the Fermi golden rule for spontaneous emission.
The other term can be shown to satisfy $\Gamma\Theta(r)\Theta(t - r/c)e^{- i\delta_L r/c} \tilde{\psi}(-r,t - r/c)$, under
reasonable approximations, namely, $g_\nu \approx g_{\nu_A}$ and the continuum limit $\sum_\nu \rightarrow \int d\nu \rho_{1D}$.
The detuning is defined as $\delta_L \equiv \nu_L - \nu_A$ and the 1D density of modes is $\rho_{1D} = L/(\pi c)$.
Finally, we have been able to eliminate self-consistently the dependence on the two-photon amplitude, finding
\beq
\left[\frac{\partial}{\partial t} + c\frac{\partial}{\partial r}\right]\tilde{\psi}(r,t) =   - \frac{\Gamma}{2}\ \tilde{\psi}(r,t)
- \Gamma\  \Theta(r)\Theta(t - r/c) e^{- i\delta_L r/c} \tilde{\psi}(-r,t - r/c).
\label{psi}
\eeq

Note that in the high detuning limit, $\delta_L \gg \Gamma$, the solution is simply
$\tilde{\psi}(r,t) = e^{-\frac{\Gamma}{2}t} \tilde{\psi}(r-ct,0)$, i.e., the product of a decaying atom and
a freely propagating photon. The general solution for Eq.(\ref{psi}) reads
\begin{eqnarray}
\tilde{\psi}(r,t) &=& \tilde{\psi}(r - ct,0)e^{-\frac{\Gamma}{2}t}
-\Gamma\Theta(r)\Theta(t - r/c)e^{-\frac{\Gamma}{2}t}e^{-\left(\frac{\Gamma}{2}- i \delta_L\right)(t - r/c)}
\int_{t -  r/c}^t  e^{\left(\frac{\Gamma}{2}-i\delta_L\right)t'} \ \tilde{\psi}(-ct',0)\ dt',
\label{ph}
\end{eqnarray}
where the initial condition of the wavepacket is written in $\tilde{\psi}(r,0)$.
In the following we restrict the study to the case of an incident photon of exponential shape, as if it had been spontaneously emitted by a neighboring atom of natural linewidth $\Delta$
and central frequency $\nu_L$, i.e., $\psi(r,0) = \mathcal{N} e^{\left(\frac{\Delta}{2}+i\nu_L\right)\frac{r}{c}} \Theta(- r)$, where $\mathcal{N}^2 = 2\pi\rho_{1D}\Delta$ stands for normalization. To obtain inversion of population of a two-level system in an experiment, a pulsed-laser excitation can be used \cite{jclaudon}.
Another option is to incoherently pump the two-level atom through a third level, which is properly described by a different formalism \cite{1d}.

\subsection{Two-photon amplitude in real-space representation}
The two-photon component are also defined in real space as $\phi(r_1,r_2,t) \equiv
\sum_{\nu,\nu'} \phi_{\nu,\nu'}(t)e^{i k_\nu r_1+i k_{\nu'}r_2}$. Within the same approximations done before, we find
\beq
\phi(r_1,r_2,t) = \sqrt{\frac{\pi\rho_{1D}\Gamma}{2}} [\Theta(t- r_2/c)\Theta( r_2)\psi(r_1-r_2,t- r_2/c)
+\Theta(t- r_1/c)\Theta(r_1)\psi(r_2-r_1,t- r_1/c)]
\label{phiaa}
\eeq
written in the original reference frame.
In the limit of vanishing interaction between the incoming photon and the atom (e.g., $\delta_L \gg \Gamma$), the two-photon wavefunction can be shown to consist in a symmetrized product of two independent single-photon wavefunctions, one describing spontaneous emission ($\psi_{\mathrm{sp.em.}}(r,t) = \exp[-(\Gamma/2+i\nu_A)(t-r/c)]$) and another describing the free propagation of the field ($\psi_{{\mathrm{free\ prop.}}}(r-ct,0)$).
In the limit of $t_\infty \gg 1/\Gamma$, the excitations are entirely in the field and the dynamics also reduces to free propagation, so that one can define the function $\phi_\infty$ checking

\beq
\phi(r_1,r_2,t_\infty) = \phi_\infty(ct_\infty-r_1 , ct_\infty-r_2).
\label{f}
\eeq

\section{Time resolved signatures of stimulated emission }

\subsection{Excited state population dynamics}

\begin{figure}[h!]
\begin{center}
\includegraphics[width=0.6\linewidth]{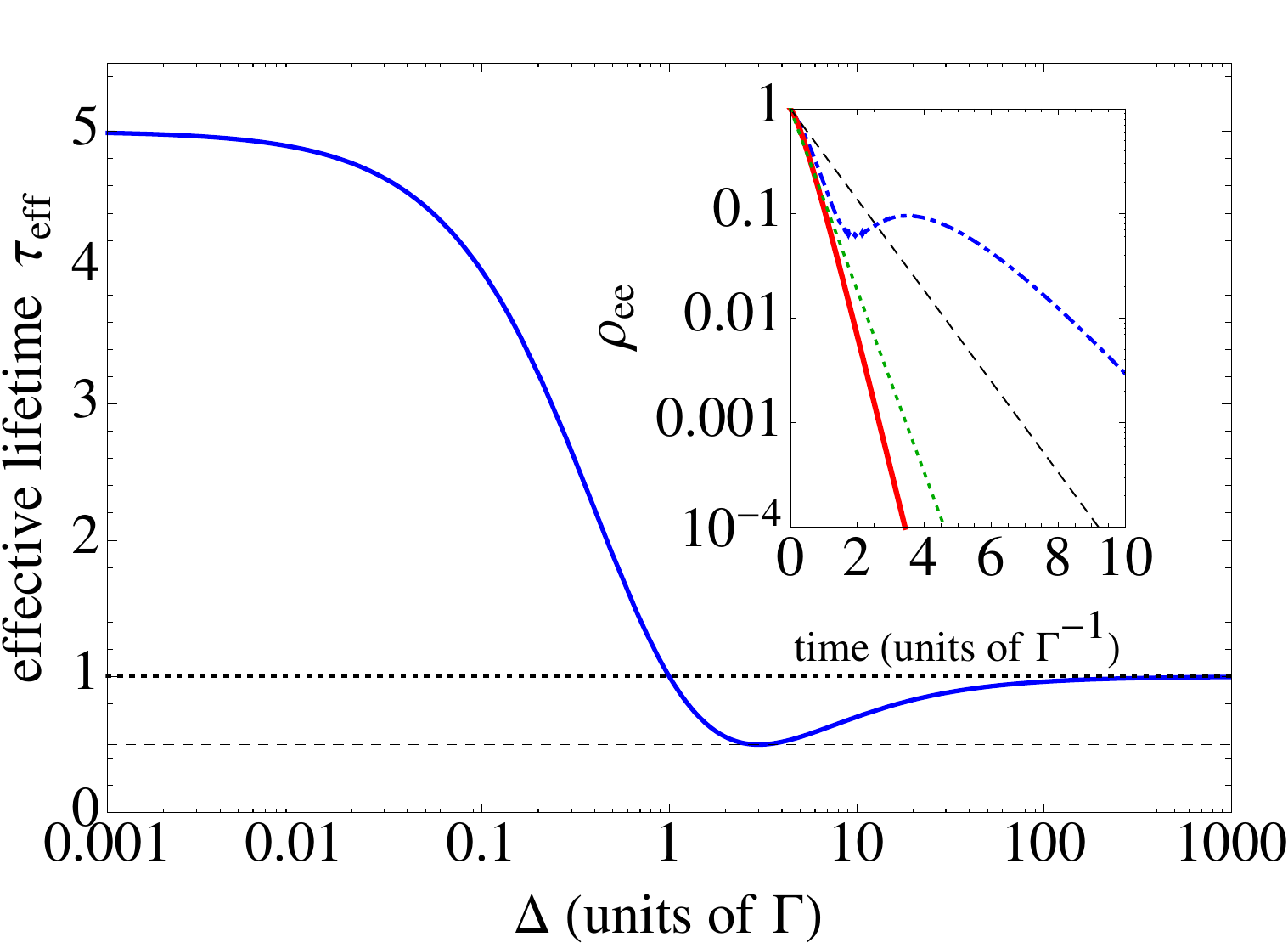}
\caption{Effective population lifetime $\tau_{\mathrm{eff}}$ as a function of $\Delta$. The dashed line indicates the optimal value $\tau_{\mathrm{opt}} = 1/2$. In the inset: excited-state population $\rho_{ee}$ decay in time. The blue dot-dashed curve, for $\Delta = \Gamma$, shows stimulation for very short times and reabsorption after $2\Gamma^{-1}$. The red solid curve represents the maximally stimulated relaxation, for $\Delta = 3\Gamma$, which for times smaller (bigger) than $\Gamma^{-1}$ goes above (below) the reference $\exp{(-2\Gamma t)}$ (green dotted). The black dashed curve represents the spontaneous emission evolution $\exp{(-\Gamma t)}$.}
\label{fig:intdecay}
\end{center}
\end{figure}

We first investigate the dynamical signature of stimulated emission and compute the excited state population
$\rho_{ee}(t) = \langle e| \mathrm{Tr}_{\mathrm{field}}[|\psi(t)\rangle\langle\psi(t)|]|e\rangle$. Given our choice for the initial state,
\begin{eqnarray}
\rho_{\mathrm{ee}}(t) = e^{-(\Gamma +\Delta )t}
\left\lbrace 1+|1+Q|^2(e^{\Delta t}-1)+|Q|^2\frac{\Delta }{\Gamma }(e^{\Gamma t}-1)
-2\mathrm{Re}\left[ (Q^*+|Q|^2)\frac{2\Delta }{\Gamma +\Delta +2i\delta }
(e^{\frac{\Gamma +\Delta +2i\delta }{2}t}-1) \right]\right\rbrace ,
\end{eqnarray}
where we have defined the factor $Q \equiv 2 \Gamma/(\Gamma-\Delta-2i\delta_L)$ and $\mathrm{Re}$ stands for the real part.
First, we note that in the strongly detuned case $\delta_L \gg \Gamma$, the atom is transparent to the incident photon, so that $\rho_{ee}(t)|_{\delta_L \rightarrow \infty} = \exp(-\Gamma t)$, which is spontaneous emission.
We then focus on the resonant case and study the influence of the width of the pulse on the dynamics of the atomic relaxation. A convenient signature is given by the adimensional effective lifetime
$\tau_{\mathrm{eff}} \equiv \Gamma \int_0^\infty \rho_{\mathrm{ee}}(t)\ dt$, which reads
\beq
\tau_{\mathrm{eff}} = 1-\frac{4\Gamma}{\Gamma+\Delta}+\frac{8\Gamma^2}{(\Gamma+\Delta)^2},
\eeq
and is plotted in Fig. \ref{fig:intdecay}.
The case $\Delta \gg \Gamma$ corresponds to a very short pulse in time and also gives rise to free propagation of the pulse followed by spontaneous atomic decay ($\tau_{\mathrm{eff}} =1$) as the spectral overlap with the atom is negligible.
On the contrary, a highly monochromatic photon ($\Delta \ll \Gamma$) increases the effective lifetime due to the efficient absorption of the incident wavepacket, after the atom has spontaneously relaxed.
Shortening of the atomic lifetime induced by stimulation can be observed for $\Gamma < \Delta \lesssim 100\Gamma$.
If $\Delta = \Gamma$, the integrated effect of the stimulation exactly compensates for the total absorption and $\tau_{\mathrm{eff}} = 1$.  The corresponding dynamics is plotted in the inset of Fig.\ref{fig:intdecay}, where a fast decay is followed by a re-excitation induced by the absorption of the tail of the photonic wave packet.
The case where $\Delta = 3 \Gamma$ gives the shortest effective atomic lifetime, namely half the spontaneous emission lifetime. In this situation, a fast population decay induced by stimulated emission takes place, minimizing the reabsorption effect and leading to optimal irreversible stimulated emission.
Note that for times $t<\Gamma^{-1}$ (resp. $t>\Gamma^{-1}$) the population is bigger (resp. smaller) than the reference $\exp{(-2\Gamma t)}$ so
that the effective lifetimes of both curves are equal.

At this point we mention that an intuitive derivation of the optimal effective lifetime can be obtained by modelling the 1D atom as an ultimate gain medium, e.g. a single emitter initially inverted, irreversibly decaying into a collection of modes $\left\{n_j\right\}$. Einstein rate equations for the excited-state population can be written \cite{Einstein,siegman}
\beq
\frac{d}{dt} \rho_{\mathrm{ee}} \stackrel{?}{=} -\Gamma (1-\beta) (1+n_l)\ \rho_{\mathrm{ee}} - \Gamma (1+n_k) \beta\ \rho_{\mathrm{ee}},
\nonumber\eeq
where $n_k$ is the number of photons in the stimulating mode, $n_l$ the number of photons in the other modes. $\beta$ is the fraction of coupling with the one-dimensional (1D) channel with respect to the 3D continuum of modes \cite{1d}. Let us assume that $n_l=0$. In the conventional 3D scenario, $\beta \ll 1$, the emitter's decay is not modified unless $n_k > \beta^{-1}$. In the 1D case under consideration, $\beta = 1$, a single photon ($n_k = 1$) is enough to stimulate the transition and shorten the lifetime by a factor of $2$ ($\dot{\rho}_{\mathrm{ee}} = -2\Gamma \rho_{\mathrm{ee}}$).

Even though it is intuitive, this simplified picture does not capture all the physics of the problem. The atomic evolution presented here does not obey a rate equation, justifying the Hamiltonian resolution adopted here. To give an example, let us consider the best stimulation condition, $\Delta = 3\Gamma$. In that case,
\begin{eqnarray}
\rho_{\mathrm{ee}} &=&  -2 e^{-4\Gamma t} + 3 e^{-3\Gamma t},
\end{eqnarray}
so
\begin{eqnarray}
\frac{d\rho_{\mathrm{ee}}}{dt} = 8\Gamma e^{-4\Gamma t} - 9 \Gamma e^{-3\Gamma t}
\neq -2\Gamma\ \rho_{\mathrm{ee}},
\end{eqnarray}
clearly evidencing a difference between standard laser systems and the present scenario.

Emitters weakly coupled to directional leaky cavities are often suggested as potential one-dimensional atoms \cite{1D,GNL}. It is worth noticing that atomic emission cannot be stimulated by an additional photon initially prepared in a dissipative monomode cavity. As a matter of fact, this photon would escape the cavity in a typical time $1/\kappa$, way too fast to stimulate any atomic emission that would take place on $\kappa/g^2$, where the atom-cavity coupling strength $g$ checks
in the weak coupling regime $g\ll \kappa$. Hence, stimulated emission by a single photon can only be simultaneously optimal and irreversible with pulse-shaped photons propagating in broadband waveguides.

\subsection{Temporal correlations in the output field}

A consequence of the atomic relaxation enhancement is the emergence of bunching in the output field. This kind of photonic temporal correlation finds application, for instance, in quantum lithography \cite{kok,Shih}. Photodetection signals are registered with a detector positioned at $r_d \gg c/\Gamma$ as pictured in Fig.\ref{fig:1datom},
so that all the excitations are in the light field. This regime corresponds to free propagation, so that the characteristics of the field only depend on the variable $r-ct$ and eq.(\ref{f}) is valid. We shall use the reference frame of the photodetector, of origin $r_d$ and $t_d=r_d/c$. The density of probability to detect one click at time $t$ and one click at time $t+\tau$ is obtained from the second-order correlation function \cite{glauber} that in our case checks $G^{(2)}(t,t+\tau) = |\phi_\infty (ct,c(t+\tau))|^2$.
From Eqs.(\ref{ph}) and (\ref{phiaa}), we find for $\tau > 0$
\beq
G^{(2)}(t,t+\tau) = \Delta\Gamma e^{-(\Gamma+\Delta)t}
\left|
(1+Q) e^{-\frac{\Gamma}{2}\tau}+
(1-Q) e^{-\left(\frac{\Delta}{2}+i\delta_L\right)\tau}
\right|^2,
\eeq
where $Q = 2\Gamma/(\Gamma-\Delta-2i\delta_L)$ is the effective coupling factor as defined in the atomic population dynamics. A clear interpretation can be given to the expression above. Between time $t$ and time $t+\tau$, the system is projected on the single excitation subspace, giving rise to two possible situations. In one case, the first click comes from the incident field. The remaining excitation is in the atom, that will finally spontaneously relax: this corresponds to the dynamics $\exp{[-\Gamma \tau/2]}$ of weight $1+Q$. In the other case, the photon emitted by the atom clicks before the incident one. This second situation, that corresponds to stimulated emission, gives rise to the component $\exp{[-(\Delta/2+i\delta_L)\tau]}$ weighted by $1-Q$. The condition $1+Q = 0$ allows to cancel spontaneous emission and to maximize stimulated emission. This is obtained for $\Delta = 3\Gamma$ and $\delta_L = 0$, namely the very same condition that minimizes the atomic lifetime.

\begin{figure}[h!]
\begin{center}
\includegraphics[width=0.6\linewidth]{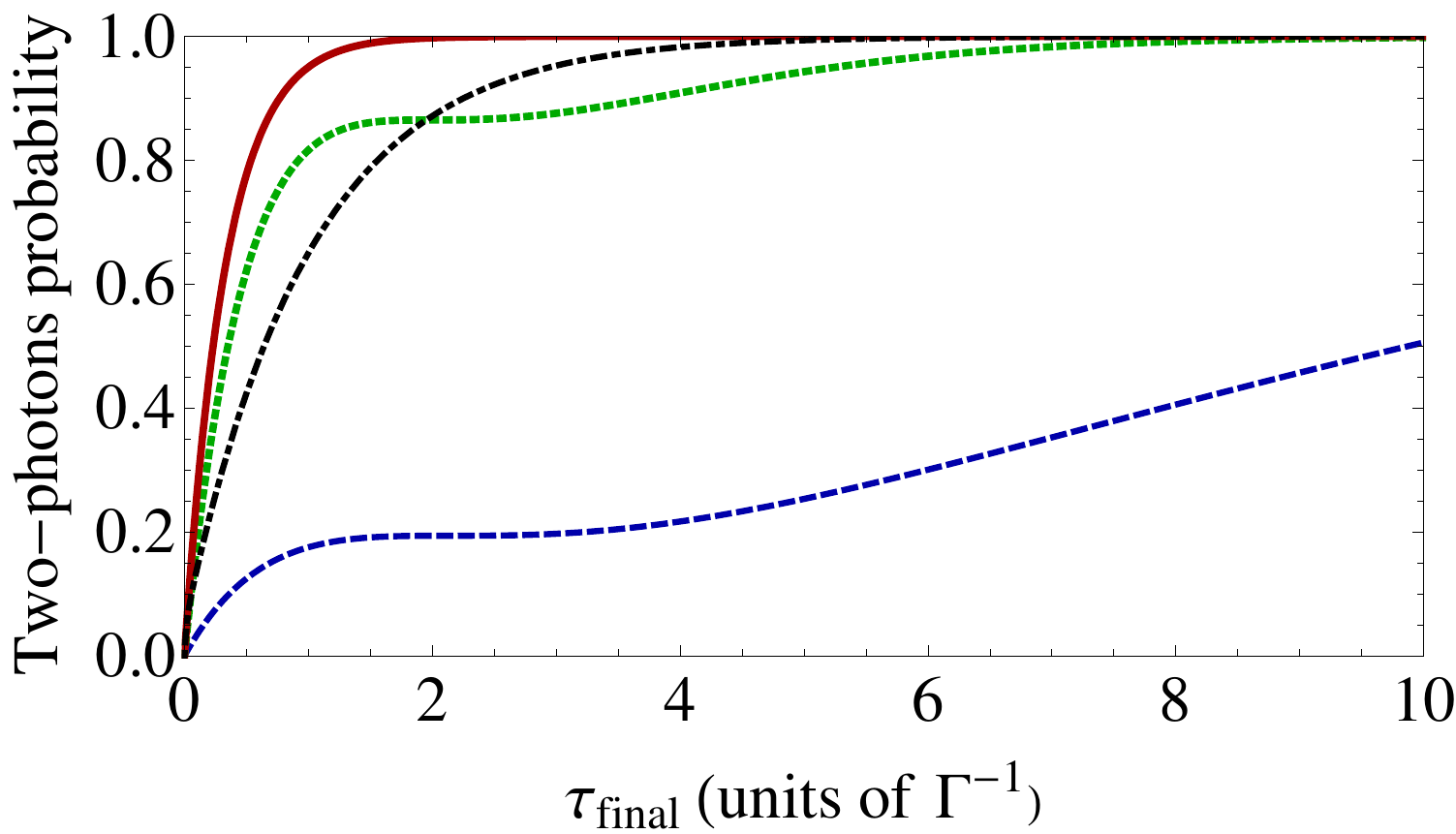}
\caption{Probability of detecting 2 photons within a time interval $\tau_{\mathrm{final}}$ as a function of $\tau_{\mathrm{final}}$ for $\Delta = 0.1 \Gamma$ (blue), $1\Gamma$ (green), $3\Gamma$ (red), and $100\Gamma$(black).}
\label{fig:Delta}
\end{center}
\end{figure}

\begin{figure}[h!]
\begin{center}
\includegraphics[width=0.6\linewidth]{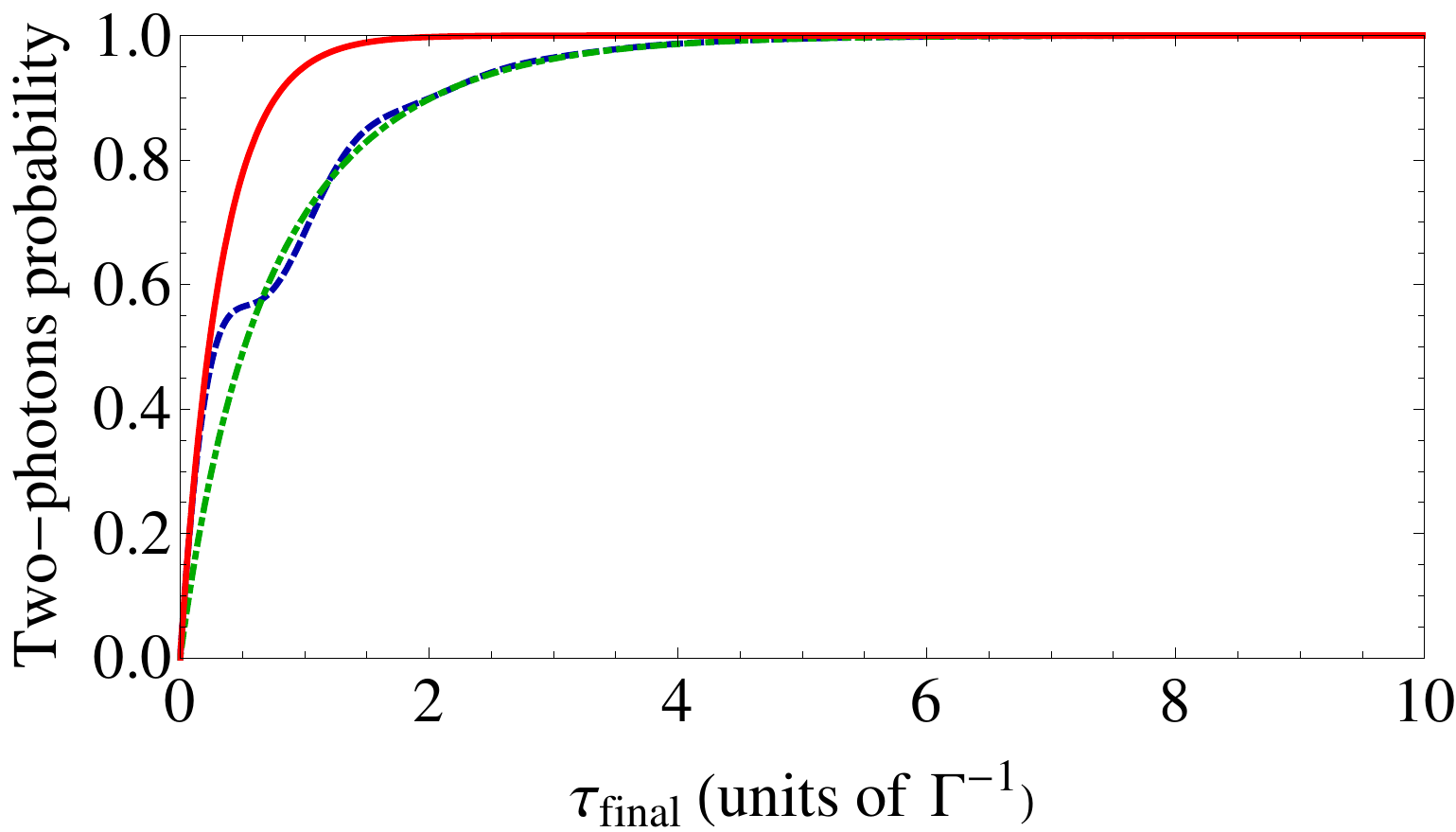}
\caption{Probability of detecting 2 photons within a time interval $\tau_{\mathrm{final}}$ as a function of $\tau_{\mathrm{final}}$ for an incident photon of different detunings: $\delta_L = 0$ (red), $\delta_L =5\Gamma$ (blue), and $\delta_L\rightarrow \infty$ (green).}
\label{fig:delta}
\end{center}
\end{figure}

This optimal regime for stimulated emission leads to the emission of both photons in a typical time $1/3\Gamma$, giving rise to bunching in the output field. The effect can be observed on Fig.\ref{fig:Delta},
where we have plotted the probability of detecting the two photons within a time $\tau_{\mathrm{final}}$ ($\int_0^{\tau_{\mathrm{final}}}d\tau \int_0^\infty dt G^{(2)}(t,t+\tau)$) as a function of $\tau_{\mathrm{final}}$, at resonance, for different values of the parameter $\Delta$. The fastest convergence is obtained for $\Delta = 3\Gamma$.
Keeping this optimal value of $\Delta$ we have plotted the same function, on Fig. \ref{fig:delta}, for different detunings, clearly showing as well the importance of the resonance on the stimulation efficiency.

\section{Potential of 1D atoms for classical and quantum amplification}

\begin{figure}[h!]
\begin{center}
\includegraphics[width=0.6\linewidth]{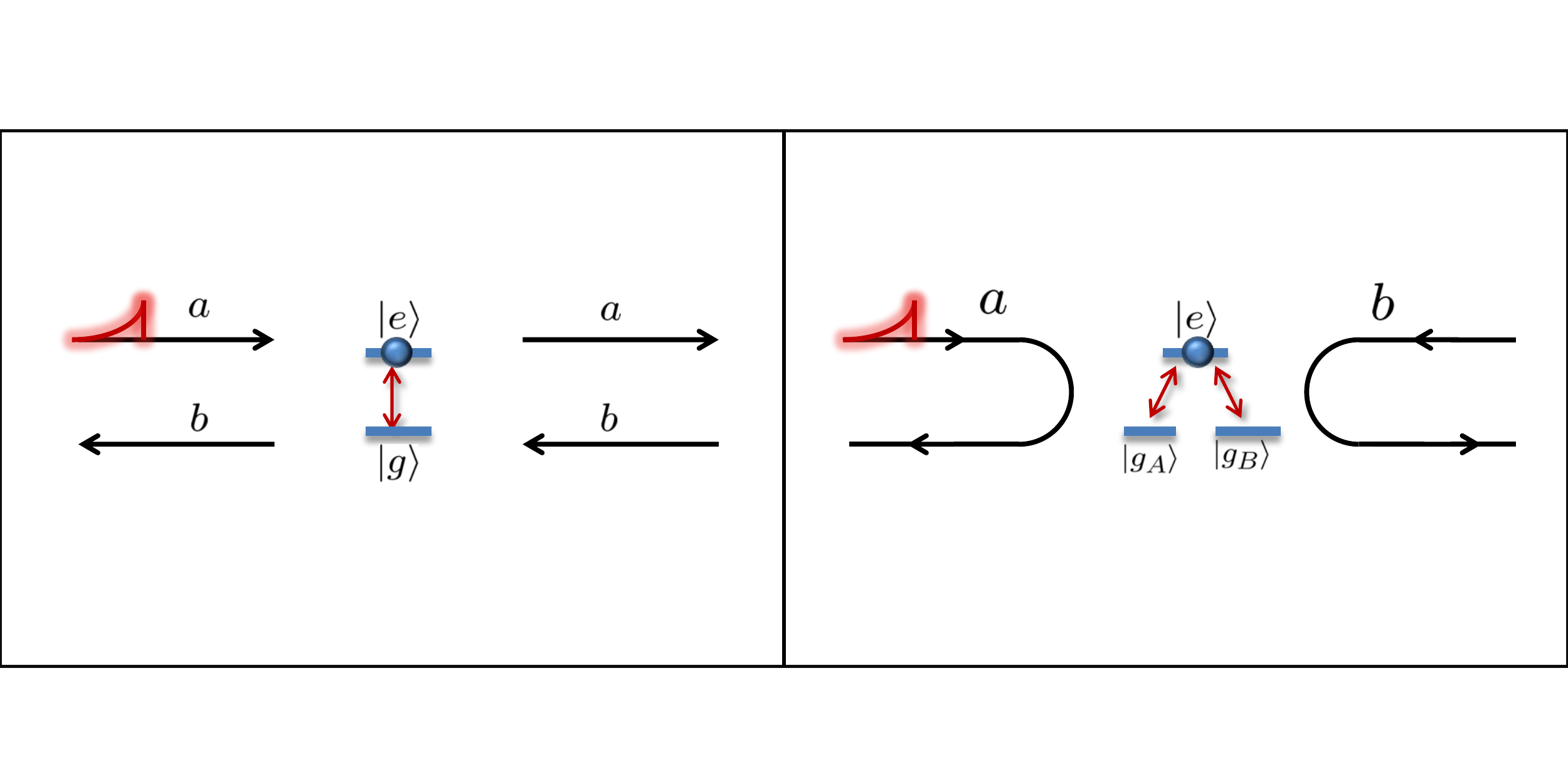}
\caption{Comparison between different types of two-continua of modes in 1D atoms. On the left, the reflection $b$/transmission $a$ are the two degrees of freedom. On the right, the degrees of freedom are represented by the $a$/$b$ polarizations. The decay of the excited-state population is the same in both cases. In contrast, the probabilities of emission in each channel, $p_{ab}$ and $p_{bb}$, crucially differ.}
\label{fig:2_1Datoms}
\end{center}
\end{figure}

Interesting applications of stimulated emission rely on the preferential emission of light in the stimulated mode rather than in other empty modes. With this aim, we revisit in this section the case of an atom interacting with two one-dimensional electromagnetic fields $a$ and $b$, one containing a single propagating photon, and the second one in the vacuum state. Two paradigmatic systems are considered in the light of the time-resolved study performed above (see Fig.\ref{fig:2_1Datoms}): a two-level atom in a transmitting/reflecting infinite waveguide \cite{ahlala}, and a lambda shaped three level atom in a semi-infinite waveguide closed by a mirror \cite{clone}. The fields $a$ and $b$ correspond respectively to photons propagating to the right and to the left, or photons of two orthogonal polarizations in the half waveguide.

\subsection{Models and atomic population dynamics}

\subsubsection{A two-level atom in a transmitting waveguide}

The Hamiltonian for such system reads $H = H_{\mathrm{1D}}+H_{\mathrm{atom}}+H_{\mathrm{int}}$ and the new interaction is given by
\beq
H_{\mathrm{int}} = -i\sum_{\nu=0}^\infty \hbar g_\nu \left[(a_\nu+b_\nu)\sigma_+ - \mathrm{H.c.} \right],
\eeq
where $a_\nu$ is the forward and $b_{\nu}$ is the backward propagating modes \cite{domokos}. Following the formalism in Ref.\cite{ahlala},
one can decompose the problem in two parts by using the even $\hat{e}_\nu = (a_\nu +  b_\nu)/\sqrt{2}$ and odd $\hat{o}_\nu = (a_\nu - b_\nu)/\sqrt{2}$
field modes. The new interaction Hamiltonian depends only on the even modes,
\beq
H^{e}_{\mathrm{int}} = -i\sum_{\nu=0}^\infty \hbar \sqrt{2} g_\nu \left[\hat{e}_\nu \sigma_+ - \mathrm{H.c.} \right],
\label{full1D}\eeq
while the odd modes contribute only to the free evolution. The spontaneous emission rate $\Gamma'$ of the transmitting 1D atom is doubled with respect to the case of an atom in a closed waveguide, i.e.  $\Gamma' = 2\Gamma$, which is due to the doubling of the resonant modes available for atomic relaxation.

In the case of stimulated emission under study, the initial state of the field writes $\ket{1_a, 0_b}$, which
checks $\ket{1_a,0_b} = (\ket{1_o,0_e}+\ket{0_o,1_e})/\sqrt{2}$, where $\ket{1_o}$ (resp. $\ket{1_e}$) is the corresponding propagating photon in the odd (resp. even) mode.
Hence, the system can follow two equiprobable paths, one giving rise to stimulated (resp. spontaneous) emission, so that the excited state population of the full transmitting/reflecting 1D atom checks
\beq
\rho^{\mathrm{full}}_{\mathrm{ee}}(t) = \frac{1}{2}\ \rho^{\mathrm{semi}}_{\mathrm{ee}}(t) + \frac{1}{2}\  e^{-\Gamma' t}, \nonumber
\eeq
in terms of the population $\rho^{\mathrm{semi}}_{\mathrm{ee}}(t)$ of the two-level 1D atom closed by one mirror.

As previously, the minimal value of the effective atomic lifetime for this system is obtained at $\Delta = 3\Gamma'$.
Because of the intrinsic presence of spontaneous emission in the odd modes, the effective lifetime is averaged and increases to $\tau_{opt} = 3/4$ as demonstrated in \cite{ahlala}.

\subsubsection{A lambda shaped 1D atom}

In the case of a three level atom in the lambda configuration coupled to a half waveguide, the initially excited
atom $\ket{e}$ can decay by emitting an $a$ or a $b$ polarized photon, ending up in state $\ket{g_a}$ or $\ket{g_b}$ respectively. The interaction Hamiltonian writes now

\beq
H_{\mathrm{int}} = -i\sum_{\nu=0}^\infty \hbar g_\nu \left[(a_\nu \sigma_a^\dagger +b_\nu \sigma_b^\dagger)- \mathrm{H.c.} \right],
\eeq
where $\sigma_a = \ket{g_a}\bra{e} $ and  $\sigma_b = \ket{g_b}\bra{e} $ are the lowering atomic operators. As above, the spontaneous emission rate is doubled, and the system
evolution in the presence of a single propagating photon in mode $a$ also splits into two paths, namely spontaneous emission in mode $b$, or stimulated emission in mode $a$.
This formal analogy leads to the same conditions of minimization of the effective atomic lifetime.
The crucial difference between the two setups (transmitting waveguide and lambda atom) appears in the characteristics of the fields, as clarified in the next section.

\subsection{Optimal emission in the stimulated mode -- application to amplification}

To study the emission in the stimulated mode, a convenient quantity is
the density of probability $G^{(2)}_{aa}(t,t+\tau)$ of detecting a click at time $t$ and $t+\tau$ in that mode, that is given in both systems by
\beq
G^{(2)}_{\mathrm{aa}}(t,t+\tau) = \Delta\Gamma' e^{-(\Gamma'+\Delta)t} \left| (1+Q_f)\ e^{-\frac{\Gamma'}{2} \tau} + (1-Q_f)\ e^{-\left(\frac{\Delta}{2}+i\delta_L\right)\tau}\right| ^2,
\eeq
where $Q_{f} = \Gamma'/(\Gamma'-\Delta-2i\delta_{L}) = Q/2$ is the new effective coupling factor. Again, the term evolving like $e^{-\Gamma'\tau/2}$ corresponds to the case where
the first click comes from the field, and is followed by a spontaneous emission of the atom in mode $a$. The second term $e^{-\left(\frac{\Delta}{2}+i\delta_L\right)\tau}$
is due to stimulated emission of the atom in mode $a$, followed by a second click coming from the field. Extinction of spontaneous emission is realized when $1+Q_f=0$. This yields
$\Delta_{opt}=2\Gamma'$, a condition which does not minimize the effective lifetime, as it was the case in the half waveguide case.

A consequence of the optimal correlation is found in the probabilities of photon emission.
By suppressing the spontaneous emission path, one avoids the atomic emission in the mode $b$ once a first photon $a$ has been detected, strongly inducing the emission of light in mode $a$. Naturally, this condition also maximizes the probability to find two photons in the mode $a$ in the end of the interaction, which reaches $p_{aa}=2/3$ \cite{ahlala,clone}. This is the maximal probability to clone the state $a$. In the case of the lambda atom, this property can be used to clone the quantum state of the incoming photon \cite{clone}.  A convenient figure of merit for this device is the fidelity ${\cal F} = p_{aa}+p_{ab}/2$, which, in the optimal point, exactly reaches $5/6$, namely, the optimal bound for quantum cloning. Because of the rotational invariance of the global system, the cloning is universal, a highly desirable property for state estimation and quantum cryptography.

\begin{figure}[h!]
\begin{center}
\includegraphics[width=0.7\linewidth]{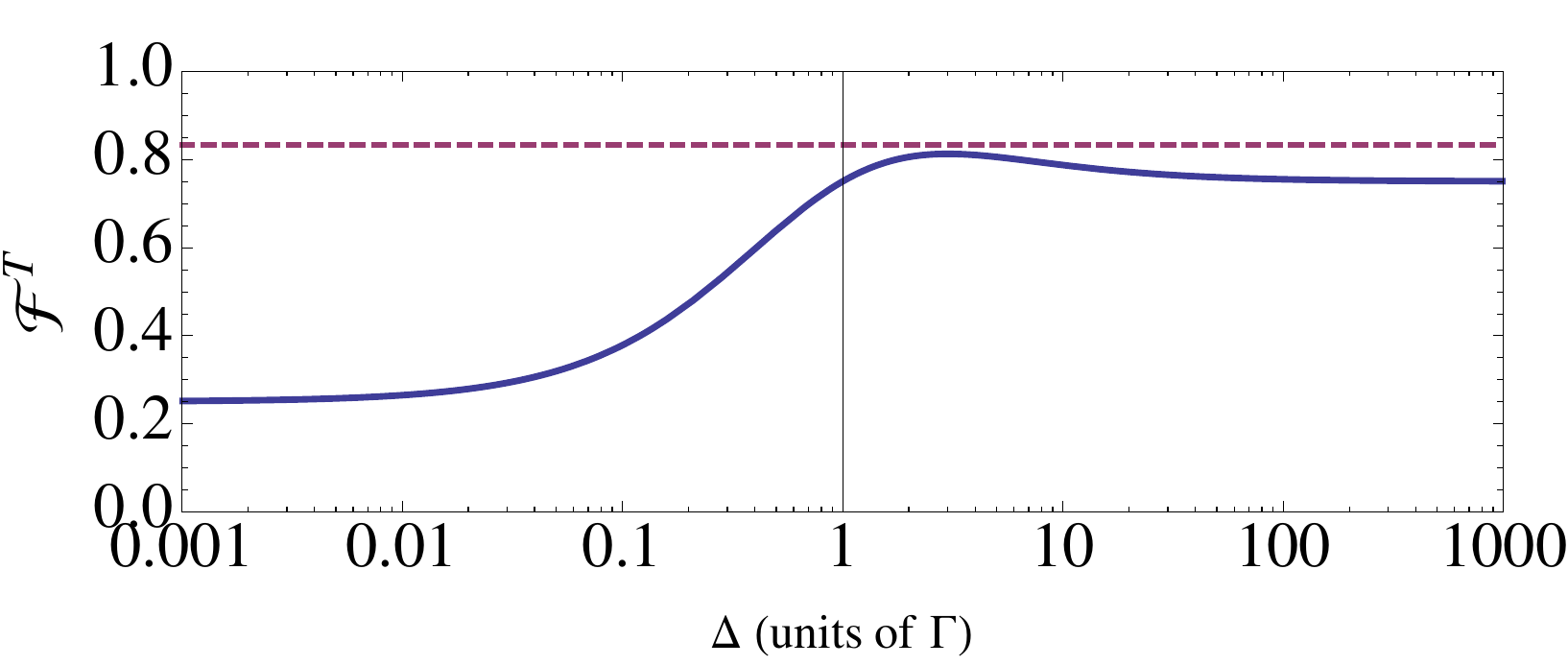}
\caption{Transmission fidelity ${\cal F}^{T}$ as a function of the packet width $\Delta$. The optimal value is ${\cal F}^{T}_{opt} = 97,5\%\times  (5/6)$, found for $\Delta = 3\Gamma'$. The dashed line indicates the maximal value ${\cal F}^{T}_{max} = 5/6$.}
\label{fig:NT}
\end{center}
\end{figure}

On the other hand, the transmitting atom can be used as an ultimate gain medium to efficiently amplify the classical state of the photon, encoded in the direction of propagation.
The convenient criterium in that case is the amplification ratio (called {\it visibility} in \cite{hwang} and {\it gain} in \cite{astafiev})
$\mathcal{A} \equiv [\mathcal{F}^{T}_{opt}-\mathcal{F}^{T}(\delta_L \rightarrow \infty)]/\mathcal{F}^{T} (\delta_L \rightarrow \infty)$, where the parameter $\mathcal{F}^{T}$ now represents the transmission fidelity of the system. Remembering that $\mathcal{F}_{max}=5/6$, and $\mathcal{F}^{T} (\delta_L \rightarrow \infty)=3/4$, one simply finds the maximal amplification ratio $\mathcal{A}_{max}=1/9$.
Contrary to the case of the lambda atom, this upper bound is not reached here. The transmission fidelity is plotted in fig.\ref{fig:NT} as a function of the packet linewidth, its maximal value being ${\cal F}^{T}_{opt} = 97,5\%\times  (5/6)$.
This difference comes from the fact that after having spontaneously emitted a photon in mode $b$, the atom in state $g$ can still interact with the incoming field in mode $a$. Thus, it is possible to finally get two photons in mode $b$,
so that $p_{bb}\neq 0$, lowering the fidelity. This situation is forbidden for a lambda atom that relaxes in state $g_b$ and becomes transparent to the incoming field.
We find $\mathcal{A} = 1/12 \approx 8,3 \%$, that means almost one order of magnitude higher than previously reported
gains, around $\sim 1\%$ \cite{hwang, astafiev}, working in the regime where the probe is continuous \cite{1d}.

In a realistic scenario, two noise sources must be taken into account, namely, the decay rate into the environmental 3D channels $\gamma$ and
the pure dephasing rate $\gamma^*$ present in solid-state systems.
The former is usually quantified by the parameter $\beta = \Gamma/(\Gamma+\gamma)$ which can reach $0.98$ in 1D nanophotonic systems made of photonic wires \cite{jclaudon} or 1D waveguides in photonic crytals \cite{photonicCrystal1,photonicCrystal2}, and almost $1$  in circuit QED \cite{Abdumalikov}.
Pure dephasing rates of $\gamma^* \approx 0.1 \Gamma$ have been measured in quantum dots \cite{dephasingQD} and superconducting qubits \cite{astafiev1D}.
The stimulation depends on the coherence of the atom-field interaction, which
can be shown \cite{1d} to depend on $\beta$ and $\gamma^*$ roughly as $\sim \beta (1-\gamma^*/\Gamma)$, at $\beta \approx 1$ and $\gamma^* \ll \Gamma$.
One can thus define a factor of trust $f_T \approx \beta(1-\gamma^*/\Gamma)$ that estimates how close to the optimal values the realistic fidelity can be. For the above mentioned systems, it can be expected of the order of $f_T \sim 90\%$. In addition to building cleaner systems, dynamical decoupling approaches \cite{plenio} can be used to reduce dephasing.

\section{Conclusions and Perspectives}
\hspace{0.3cm}We have shown the influence of the incoming photon on the atom decay as a function of the packet shape.
An irreversible and maximally accelerated stimulated emission occurs for the broadband mode-matching condition where the incoming photon is three times shorter than the spontaneously emitted one. We have also studied the influence of stimulation on the two-photon correlation function, which shows optimal photon bunching. Finally, we added a second one-dimensional field to explore quantum and classical amplification.
In the quantum case, the possibility to achieve universal optimal cloning of polarization has been presented.
In the classical case, amplification has been shown in the average transmitted field, which reaches $97.5\%$ of the ideal case. This effect has led to a transistor-like amplification that can overcome the continuous-wave approach by a factor 8. The photonic propagation makes the reported effects specially attractive as far as realistic implementations of quantum information processing are concerned.

\section*{Acknowledgements}
\hspace{0.3cm} This work was supported by the Nanosciences Foundation, Grenoble, by CNPq and Fapemig,
Brazil, by the ANR, France, through the projects WIFO and CAFE and by the Centre for
Quantum Technologies, Singapore.

\end{document}